\documentclass[aps,12pt,nofootinbib,preprint]{revtex4}
\pdfoutput=1
\usepackage{textcomp}
\usepackage{amssymb}
\usepackage{anysize}
\usepackage{bold-extra}
\usepackage{enumerate}
\usepackage{graphicx}
\usepackage{lscape}
\usepackage{mathrsfs}
\usepackage{multirow}
\usepackage{hhline}
\usepackage{makecell}
\usepackage[final]{pdfpages}
\usepackage{rotating}
\usepackage{subfig}
\usepackage{url}
\usepackage{wrapfig}
\usepackage{amsmath}
\usepackage{color}
\usepackage{fancyhdr}
\usepackage{verbatim}
\usepackage{epstopdf}
\usepackage[section]{placeins}
\usepackage[normalem]{ulem} 
\usepackage[colorlinks=true,
            linkcolor=red,
            urlcolor=blue,
            citecolor=blue]{hyperref}
\def\nn{\nonumber}
\def\bea{\begin{eqnarray}}
\def\eea{\end{eqnarray}}
\def\ba{\begin{eqnarray}}
\def\ea{\end{eqnarray}}
\def\be{\begin{equation}}
\def\ee{\end{equation}}
\def\beq{\begin{equation}}
\def\eeq{\end{equation}}

\newcommand{\tf}{\texorpdfstring}

\newcommand{\gev}{~\text{GeV}}

\newcommand{\abi}{~\text{ab}^{-1}}
\newcommand{\fbi}{~\text{fb}^{-1}}

\begin{document}

\title{Search for a heavy dark photon\\
at future $e^+e^-$ colliders}

\author{Min He$^1$\footnote{Electronic address: hemind@sjtu.edu.cn}, Xiao-Gang He$^{1, 2,3}$\footnote{Electronic address: hexg@phys.ntu.edu.tw}, Cheng-Kai Huang$^{2}$\footnote{Electronic address: r01222045@ntu.edu.tw }, Gang Li$^{2}$\footnote{Electronic address: gangli@phys.ntu.edu.tw }}
\affiliation{$^1$T-D. Lee Institute and School of Physics and Astronomy, Shanghai Jiao Tong
University, 800 Dongchuan Road, Shanghai 200240\\
$^2$Department of Physics, National Taiwan University, Taipei 106\\
$^3$National Center for Theoretical Sciences, Hsinchu 300}


\begin{abstract}
A coupling of a dark photon $A'$ from a $U(1)_{A'}$ with the standard model (SM) particles can be generated through kinetic mixing represented by a parameter $\epsilon$. A non-zero $\epsilon$ also induces a mixing between $A'$ and $Z$ if dark photon mass $m_{A'}$ is not zero. This mixing can be large when $m_{A'}$ is close to $m_Z$ even if the parameter $\epsilon$ is small. Many efforts have been made to constrain the parameter $\epsilon$ for a low dark photon mass $m_{A'}$ compared with the $Z$ boson mass $m_Z$. We study the search for dark photon in $e^+e^- \to \gamma A' \to \gamma \mu^+ \mu^-$ for a dark photon mass $m_{A'}$ as large as kinematically allowed at future $e^+e^-$ colliders. For large $m_{A'}$, care should be taken to properly treat possible large mixing between $A'$ and $Z$. We obtain sensitivities to the parameter $\epsilon$ for a wide range of dark photon mass at planed $e^+\;e^-$ colliders, such as Circular Electron Positron Collider (CEPC),  International Linear Collider (ILC) and Future Circular Collider (FCC-ee). For the dark photon mass $20~\text{GeV}\lesssim m_{A^{\prime}}\lesssim 330~\text{GeV}$, the $2\sigma$ exclusion limits on the mixing parameter are $\epsilon\lesssim 10^{-3}-10^{-2}$. The CEPC with $\sqrt{s}=240~\text{GeV}$ and FCC-ee with $\sqrt{s}=160~\text{GeV}$ are more sensitive than the constraint from current LHCb measurement once the dark photon mass $m_{A^{\prime}}\gtrsim 50~\text{GeV}$. For $m_{A^{\prime}}\gtrsim 220~\text{GeV}$, the sensitivity at the FCC-ee with $\sqrt{s}=350~\text{GeV}$ and $1.5~\text{ab}^{-1}$ is better than that at the 13~TeV LHC with $300~\text{fb}^{-1}$, while the sensitivity at the CEPC with $\sqrt{s}=240~\text{GeV}$ and $5~\text{ab}^{-1}$ can be even better than that at 13~TeV LHC with $3~\text{ab}^{-1}$ for $m_{A^{\prime}}\gtrsim 180~\text{GeV}$. 
\end{abstract}

\maketitle



\section{Introduction}
\label{sec:intro}
If there is an additional $U(1)_{A'}$ symmetry beyond the standard model (SM) gauge symmetry $SU(3)\times SU(2)_L\times U(1)_Y$, 
a non-zero coupling to the gauge particle $A'$ of $U(1)_{A'}$ can be generated due to kinetic mixing between the gauge field of $U(1)_{A'}$ and the SM hypercharge field at the renormalizable level~\cite{kinetic-mixing,kinetic-mixing1}. This gauge symmetry $U(1)_{A'}$ is referred as a dark gauge symmetry since the SM particles have zero charge of this $U(1)_{A'}$ group and naively invisible. The kinetic-mixing induced $A'$ coupling to the SM particles is proportional to the electromagnetic coupling, therefore $A'$ is usually referred as dark photon. This is a portal between a possible dark sector and the SM sector. The existence of $U(1)_{A'}$ has many interesting effects in particle physics, astrophysics and cosmology~\cite{kinetic-mixing,kinetic-mixing1,kinetic-mixing2,kinetic-mixing3}. Great efforts have been made to search for a dark photon through various processes and stringent limits have been obtained for the kinetic mixing parameter $\epsilon$ for a given dark photon mass $m_{A'}$~\cite{limits,limits1,limits2}. There are strong constraints on $\epsilon$ for a low dark photon mass 
$ m_{A'}$ (less than 10 GeV or so) from various low energy facilities and rare decays of known particles.  There are fewer studies of constraints on dark photon with a larger mass. Experiments, such as LHCb, ATLAS, CMS and SHiP at CERN, may provide some important information~\cite{limits1,limits2}. LHCb can provide stringent constraint on the kinetic mixing for dark photon mass larger than 10 GeV~\cite{limits2,Aaij:2017rft}. It has also been shown that the ATLAS and CMS may provide even better constraint~\cite{limits2,Blinov:2017dtk} at dark photon mass around $40\gev\sim 70\gev$ by studying the Drell-Yan process $pp \to X \mu^+\mu^-$. There are several high energy $e^+e^-$ colliders in the plan~\cite{FCC-ee}. These colliders can provide constraints on the mixing parameter $\epsilon$ for a wide range of the dark photon mass. 

In this work we extend our previous study~\cite{He:2017ord}, using $e^+e^- \to \gamma \mu^+\mu^-$, to obtain constraints on $\epsilon$ for dark photon mass in the full range which can be covered by a future $e^+ e^-$ collider. Compared with $pp \to X \mu^+\mu^-$, final states in $e^+e^- \to \gamma \mu^+\mu^-$ are easier to be studied. We find that a better constraint on $\epsilon$ as a function of $m_{A'}$ may be possible at some of the planned $e^+e^-$ colliders. The same $e^+e^- \to \gamma \mu^+\mu^-$ process had been used by BaBar~\cite{babar} to set stringent constraints on the relevant parameters, but the reach of the dark photon mass is limited to be lower than 10 GeV or so. We will study the possibility to search for a heavier dark photon at future $e^+e^-$ colliders, CEPC, ILC and FCC-ee, through the process $e^+ e^-\to \gamma A^{\prime } \to \gamma \mu^+\mu^-$. There are some other studies of heavy dark photon at future $e^+e^-$ colliders~\cite{Karliner:2015tga,Liu:2017lpo}, which will be discussed later. 

In our study, we perform a detailed detector simulation with more moderate selection cuts based on a realistic muon momentum resolution.
We find that the $2\sigma$ exclusion limits on $\epsilon$ for the dark photon from 20~GeV to 330~GeV  can reach $\lesssim 10^{-3}-10^{-2}$ at future $e^+e^-$ colliders. The CEPC with $\sqrt{s}=240\gev$ and FCC-ee with $\sqrt{s}=160\gev$ are more sensitive to $\epsilon$ than the constraint from current LHCb measurement once the dark photon mass $m_{A^{\prime}}\gtrsim 50\gev$. We also obtain the constraint on $\epsilon$ for $150\gev\lesssim m_{A^{\prime}}\lesssim 350\gev$ from the direct searches in the Drell-Yan process $pp\to X\mu^+\mu^-$ using the 13 TeV LHC measurements with $\mathcal{L}=36.1\fbi$~\cite{Aaboud:2017buh} and project it to the measurements with $\mathcal{L}=300\fbi$ and $3\abi$. The corresponding constraints are $\epsilon\lesssim 8.3\times 10^{-3}$, $\epsilon\lesssim 4.8\times 10^{-3}$ and $\epsilon\lesssim 2.7\times 10^{-3}$ for $150\gev\lesssim m_{A^{\prime}}\lesssim 300\gev$, and become weaker for $m_{A^{\prime}}\gtrsim 300\gev$. For $m_{A^{\prime}}\gtrsim 220\gev$, the sensitivity at the FCC-ee with $\sqrt{s}=350\gev$ and $1.5\abi$ is better than that using $pp\to X \mu^+\mu^-$ at the 13~TeV LHC with $300\fbi$. We can also achieve a better sensitivity at the CEPC with $\sqrt{s}=240\gev$ and $5\abi$ than that at 13~TeV LHC with $3\abi$ for $m_{A^{\prime}}\gtrsim 180\gev$. 

The paper is arranged as the following. In section~\ref{sec:formalism}, we discuss the interactions between the dark photon with the SM sector, where the complete couplings of dark photon with arbitrary mass to the SM particles are studied. In section~\ref{sec:prod_decay}, we discuss the production and decays of dark photon at $e^+e^-$ colliders.  In section~\ref{sec:simulation}, a detailed collider simulation with dark photon mass ranging from 20~GeV to that as kinematically allowed at future $e^+e^-$ colliders are performed. In section~\ref{sec:summary}, we summarize our results.

\section{Couplings of dark photon to the SM particles}
\label{sec:formalism}

We now study the kinetic mixing effects on the interactions of $A'$ and $Z$ with other SM particles. For $A'$ with mass smaller than the $Z$ boson mass, the mixing effects have been studied in details. Since we will allow the $A'$ mass from small to be larger than $Z$ bosom mass, care should be taken in particular when $m_{A'}$ is very close to $m_Z$ where the mixing can be large.

A dark photon field $A_{0}'$ from an extra $U(1)_{A'}$ gauge group can indirectly interact through a gauge kinetic mixing term $F'_{0,\mu\nu} B_{0}^{\mu\nu}$ with the SM sector. Here $F'_{0,\mu\nu}=\partial_\mu A^{\prime}_{0\nu}-\partial_{\nu} A^{\prime}_{0\mu}$ and $B_{0,\mu\nu} = \partial_\mu B_{0\nu}-\partial_\nu B_{0\mu}$, and $A'_{0}$ and $B_{0}$ are the $U(1)_{A'}$ and $U(1)_Y$ gauge fields, respectively. It is interesting to note that this term should naturally exist since there is no symmetry to prevent it to appear in the relevant Lagrangian even one requires renormalizability. With gauge kinetic mixing, the renormalizable terms involving these two $U(1)$ gauge fields are given by~\cite{kinetic-mixing}
\begin{eqnarray}
L_{\mbox{kinetic}} = -{1\over 4} B_{0}^{\mu\nu}B_{0,\mu\nu} -{1\over 2} \sigma F'_{0, \mu\nu}B_0^{\mu\nu} - {1\over 4} F'_{0,\mu\nu}F_0^{\prime \mu\nu}\;.
\end{eqnarray}
The $U(1)_Y$ gauge field $B_0$ is a linear combination of the photon $A_0$ and the $Z_0$ boson fields, $B_0 = c_W A_0 - s_W Z_0$ with $c_W = \cos\theta_W$ and $s_W = \sin\theta_W$, where $\theta_W$ is the weak interaction Weinberg angle.  

To make the above Lagrangian in the canonical form, that is, there is no crossing term, one needs to redefine the fields. Letting the redefined fields to be $\tilde A$, $\tilde Z$ and $\tilde A'$, we have~\cite{kinetic-mixing}
\begin{eqnarray}
\left ( \begin{array}{c}
A_0\\
Z_0\\
A'_0
\end{array}
\right ) = \left ( \begin{array}{ccc}
 1&0&-{c_W \sigma\over \sqrt{1-\sigma^2}}\\
 0&1&{s_W \sigma\over \sqrt{1-\sigma^2}}\\
 0&0&{1\over \sqrt{1-\sigma^2}}
 \end{array}
 \right )
 \left (\begin{array}{c}
 \tilde A\\ \tilde Z\\ \tilde A'
 \end{array}
 \right )\;.
 \end{eqnarray}
 
The interaction of $\tilde A$, $\tilde Z$ and $\tilde A'$ with SM currents is given by
\begin{eqnarray}
J^\mu_{em} (\tilde A_\mu - {c_W \sigma\over \sqrt{1-\sigma^2}} \tilde A'_\mu) + J^\mu_{Z} (\tilde Z_\mu + {s_W\sigma\over \sqrt{1-\sigma^2}}  \tilde A'_\mu) +  {1\over \sqrt{1-\sigma^2}}J^\mu_{D} \tilde A'_\mu \;,\label{interaction1}
\end{eqnarray}
where $J^\mu_{em}$, $J^\mu_Z$ are the SM electromagnetic and $Z$ boson interaction currents, respectively. 
$J^\mu_D$ is the dark current in the dark sector. We will work with models where $J^\mu_D$ does not involve SM particles and assume that the width of dark photon decaying into the dark sector is zero. Therefore $J^\mu_D$ does not play a role in $e^+e^-\to \gamma \mu^+\mu^-$ and will be ignored in our later discussions.

After the electroweak symmetry breaking, the $Z_0$ boson obtains a non-zero mass $m_Z$. Depending on how the $U(1)_{A'}$ symmetry is broken, $A'_0$ boson can receive a non-zero mass which may cause a mixing with $Z_0$. If one introduces a SM singlet $S$ with a non-trivial $U(1)_{A'}$ quantum number $s_{A'}$ to break the symmetry, $A'_0$ boson
will receive  a mass $m_{A'} = g_{A'}s_{A'}v_s/\sqrt{2}$ from $(D_\mu S)^\dagger (D^\mu S)$ term in the Lagrangian. Here $g_{A'}$ is the $U(1)_{A'}$ gauge coupling constant and $v_s/\sqrt{2}$ is the vacuum expectation value $\langle S \rangle$ of $S$ field. 
In the $\tilde Z$ and $\tilde A'$ basis, they mix with each other with the mixing matrix  given by
\begin{eqnarray}
\left ( \begin{array}{cc}
 m^2_Z&{\sigma s_W\over \sqrt{1-\sigma^2}}m^2_Z\\
 {\sigma s_W\over \sqrt{1-\sigma^2}}m^2_Z&{1\over 1-\sigma^2}m^2_{A'} + {s^2_W \sigma^2\over 1-\sigma^2} m^2_Z
 \end{array}
 \right ).\label{z-a-mix}
\end{eqnarray}
The above mass matrix can be diagonalized by an unitary transformation
\begin{align}
\left( 
\begin{array}{c}
\tilde{A}\\
\tilde{Z}\\
\tilde{A}^\prime
\end{array} \right)
= \left(
\begin{array}{ccc}
1 & 0 & 0\\
0 & \frac{\sigma s_W m_Z^2}{\mathbf{M} \sqrt{1-\sigma^2}} & \frac{m_Z^2-\lambda_1}{\mathbf{M}}\\
0 & \frac{\lambda_1-m_Z^2}{\mathbf{M}} & \frac{\sigma s_W m_Z^2}{\mathbf{M} \sqrt{1-\sigma^2}}
\end{array} \right)\left(
\begin{array}{c}
A\\
Z\\
A^\prime
\end{array} \right),
\end{align}
where the normalization factor $\mathbf{M}$ is 
\begin{align}
\mathbf{M} &= \sqrt{(\lambda_1-m_Z^2)^2+\frac{\sigma^2 s_W^2}{1-\sigma^2} m_Z^4}\;,
\end{align}
and $\lambda_{1,2}$ are the eigenvalues
\begin{align}
\label{eq:eigenvalues}
\lambda_{1,2} &= \frac{1}{2}\left(m_Z^2+\frac{1}{1-\sigma^2} m_{A^\prime}^2
+ \frac{\sigma^2 s_W^2}{1-\sigma^2} m_Z^2 \pm \Delta \right),\quad \lambda_1\geq \lambda_2,\\
\Delta&\equiv \sqrt{\left(m_Z^2 - \dfrac{1}{1-\sigma^2} m_{A^\prime}^2
- \frac{\sigma^2 s_W^2}{1-\sigma^2} m_Z^2\right)^2 + \frac{4 \sigma^2 s_W^2}{1-\sigma^2} m_Z^4}.
\end{align}
For the case $m_{A^{\prime}}<m_Z$, the masses of dark photon $A^{\prime}$ and $Z$ boson are $\lambda_2$ and $\lambda_1$, respectively; while for $m_{A^{\prime}}>m_Z$, they correspond to $\lambda_1$ and $\lambda_2$. The interaction of physical dark photon with SM sector currents will be modified further compared with Eq.~\eqref{interaction1}. In the rest of this paper, we will work within this simple model for dark photon mass generation and study the consequences.

The final transformation between the basis $(A_0,Z_0,A_{0}^{\prime})^T$ and the mass eigenstate $(A,Z,A^{\prime})^T$ can be expressed as
\begin{align}
\left( 
\begin{array}{c}
A_0\\
Z_0\\
A_0^\prime
\end{array} \right)
&=V \left(
\begin{array}{c}
A\\
Z\\
A^\prime
\end{array} \right)\;,\;\;\label{eq:fullV}
V= \left(
\begin{array}{ccc}
V_{11} & V_{12}&V_{13}\\
V_{21}&V_{22}&V_{23}\\
V_{31}&V_{32}&V_{33}
\end{array} \right),
\end{align}
where the transformation matrix $V\equiv V_{-}(V_{+})$ for $m_{A^{\prime}}<m_Z (m_{A^{\prime}}>m_Z)$ are given by
\begin{align}
V_- = \left(
\begin{array}{ccc}
1 & \frac{-c_W \sigma (\lambda_1 -m_Z^2)}{\mathbf{M} \sqrt{1-\sigma^2}} & \frac{-\sigma^2 s_W c_W m_Z^2}{\mathbf{M} (1-\sigma^2)}\\
0 & \frac{s_W \sigma \lambda_1}{\mathbf{M} \sqrt{1-\sigma^2}} & \frac{1}{\mathbf{M}} \left( m_Z^2 - \lambda_1 + \frac{\sigma^2 s_W^2 m_Z^2}{1-\sigma^2} \right) \\
0 & \frac{\lambda_1 - m_Z^2}{\mathbf{M} \sqrt{1-\sigma^2}} & \frac{\sigma s_W m_Z^2}{\mathbf{M} (1-\sigma^2)}
\end{array} \right)\;,\;\;
V_+ = V_- \left(
\begin{array}{ccc}
1 & 0 & 0\\
0 & 0 & 1 \\
0 & -1 & 0
\end{array} \right).
\end{align}
After the mass diagonalization, Eq.~\eqref{interaction1} will be modified with the couplings to $J^\mu_{em}$ and $J^\mu_Z$ currents to be given by 
\begin{eqnarray}
J^\mu_{em} ( V_{11} A_\mu + V_{12} Z_\mu + \epsilon A'_\mu) + J^\mu_{Z} ( V_{22} Z_\mu + \tau A'_\mu) \;,\label{interaction2}
\end{eqnarray}
where $V_{11}=1$, $\epsilon \equiv V_{13}$ and $\tau \equiv V_{23}$.

For $m_{A^\prime} < m_Z$, we have
\begin{align}
V_{12}&=\frac{-c_W \sigma (\lambda_1 -m_Z^2)}{\mathbf{M} \sqrt{1-\sigma^2}}\;,\;\; &V_{22}&=\frac{s_W \sigma \lambda_1}{\mathbf{M} \sqrt{1-\sigma^2}}\;,\nonumber\\
\epsilon&=\frac{-\sigma^2 s_W c_W m_Z^2}{\mathbf{M} (1-\sigma^2)}\;,\;\;&\tau&=\frac{1}{\mathbf{M}} \left( m_Z^2 - \lambda_1 + \frac{\sigma^2 s_W^2 m_Z^2}{1-\sigma^2} \right),
\end{align}
while for $m_{A^\prime} > m_Z$,
\begin{align}
V_{12}&=\dfrac{\sigma^2 s_W c_W m_Z^2}{\mathbf{M} (1-\sigma^2)}\;, &V_{22} &=\dfrac{-1}{\mathbf{M}} \left( m_Z^2 - \lambda_1 + \dfrac{\sigma^2 s_W^2 m_Z^2}{1-\sigma^2} \right)\;,\nonumber\\
\epsilon &= \frac{-\sigma c_W (\lambda_1 - m_Z^2)}{\mathbf{M} \sqrt{1-\sigma^2}}\;,
&\tau &= \frac{\sigma s_W \lambda_1}{\mathbf{M} \sqrt{1-\sigma^2}}.
\end{align}

We find that if $|m_Z - m_{A^\prime}| \gg s_W m_Z \sigma$,
\begin{align}
\label{eq:mAp-mZ}
\Delta & = |m_Z^2 - m_{A^\prime}^2| + \frac{(m_{A^\prime}^4 + s_W^2 m_Z^4 - c_W^2 m_Z^2 m_{A^\prime}^2) \sigma^2}{|m_Z^2 - m_{A^\prime}^2|}+\mathcal{O}(\sigma^3),
\end{align}
and the transformation matrix $V_{\pm}$ can be expressed in an uniform form as follows:
\begin{align}
\label{eq:V case 1}
V_{+}=V_{-}&= \left(
\begin{array}{ccc}
1 & 0 & -c_W \sigma\\
0 & 1 & \dfrac{s_W \sigma m_{A^\prime}^2}{m_{A^\prime}^2 - m_Z^2}\\
0 & -\dfrac{s_W \sigma m_Z^2}{m_{A^\prime}^2 - m_Z^2} & 1
\end{array} \right)\;+\mathcal{O}(\sigma^2), 
\end{align}
which gives
\begin{align}
V_{12}=0, \quad V_{22}=1, \quad \epsilon&=-c_W\sigma,\quad \tau=\dfrac{s_{W}\sigma m_{A^{\prime}}^2}{ m_{A^{\prime}}^2-m_Z^2}.
\label{coupling}
\end{align}
It is apparent that both $\epsilon$ and $\tau$ depend on the mixing parameter $\sigma$ linearly. In this case one can express $\tau$ in terms of $\epsilon$ as 
\begin{align}
\label{coupling2}
\tau=-\dfrac{s_{W} m_{A^{\prime}}^2\epsilon}{c_W (m_{A^{\prime}}^2-m_Z^2)}.
\end{align}
We find that $\tau$ is very small if $m_{A^{\prime}}\ll m_Z$, thus it is usually neglected for light dark photon searches. For $m_{A^{\prime}}$ being close to $m_Z$, it becomes significantly large and should be taken into account.

The approximation in Eq.~\eqref{eq:mAp-mZ} is no longer valid once $|m_{A^{\prime}}- m_Z| \ll s_W\sigma m_Z$, in this case we should first take the limit of $m_{A^{\prime}}\to m_Z$ and then expand in series of $\sigma$. To the first order in $\sigma$, we obtain that
\begin{align}
\label{eq:V case 2}
V_{-} = \left(
\begin{array}{ccc}
1 & -\frac{c_W \sigma}{\sqrt{2}} & -\frac{c_W \sigma}{\sqrt{2}}\\
0 & \frac{1}{\sqrt{2}} + \frac{(3 s_W^2 - 1) \sigma}{4 \sqrt{2} s_W} & -\frac{1}{\sqrt{2}} + \frac{(3 s_W^2 - 1) \sigma}{4 \sqrt{2} s_W} \\
0 & \frac{1}{\sqrt{2}} + \frac{(s_W^2 + 1) \sigma}{4 \sqrt{2} s_W} & \frac{1}{\sqrt{2}} - \frac{(s_W^2 + 1) \sigma}{4 \sqrt{2} s_W}
\end{array} \right)+\mathcal{O}(\sigma^2)\;,\;\;
V_+ = V_- \left(
\begin{array}{ccc}
1 & 0 & 0\\
0 & 0 & 1 \\
0 & -1 & 0
\end{array} \right)
\end{align}
and the coupling constants are
\begin{align}
\epsilon&=-\dfrac{c_W\sigma}{\sqrt{2}},\quad \tau=-\dfrac{1}{\sqrt{2}}+\dfrac{(3s_W^2-1)\sigma}{4\sqrt{2}s_W},\quad \text{for}\quad m_{A^{\prime}}<m_Z,\\
\epsilon&=-\dfrac{c_W\sigma}{\sqrt{2}},\quad \tau=\dfrac{1}{\sqrt{2}}+\dfrac{(3s_W^2-1)\sigma}{4\sqrt{2}s_W},\quad\;\;\; \text{for}\quad m_{A^{\prime}}>m_Z.
\end{align}

From the above we see that the mixing between $A^{\prime\mu}_0$ and $Z_0^{\mu}$ are nearly maximal. There is a discontinuous behavior of $\tau$ for $m_{A^{\prime}}$ below and above $m_Z$.

Physically, at $m_{A'} = m_Z$ and $\sigma$ approaching zero, the mixing parameter should be zero. At that point, one should take the average of the mixing effect of $m_{A'}$ below and above $m_Z$. Then the correct limit for $m_{A'} = m_Z$ and $\sigma =0$ can be obtained.

For the value $\sigma\sim 10^{-3}-10^{-2}$ that we are interested in, one obtains that $s_W\sigma m_Z\sim 0.043\gev-0.43\gev$. 
Our detector simulation to be discussed later shows that a mass window cut  $\Delta m_{\mu^+\mu^-}<0.5\gev\sim 1.5\gev$ is appropriate for the dark photon searches in $e^+e^-\to \gamma A^{\prime}\to \gamma\mu^+\mu^-$ at future $e^+e^-$ colliders. Thus we can work with the assumption that $|m_Z - m_{A^\prime}| \gg s_W m_Z \sigma$ in section~\ref{sec:simulation}.

It is important to emphasize that $m_{A^{\prime}}$ and $m_Z$ are not the physical masses but the mass parameters of $A_{0}^{\prime}$ and $Z_0$. From Eq.~\eqref{eq:eigenvalues}, the masses for $Z$ and $A^\prime$ can be expressed as
\begin{align}
(m_{Z}^{\text{phys.}})^2 &= m_{Z}^2 + \frac{m_{Z}^4 s_W^2 \sigma^2}{m_{Z}^2 - m_{A'}^2} + \mathcal{O}(\sigma^3),\nonumber\\
(m_{A^{\prime}}^{\text{phys.}})^2 &= m_{A'}^2 + \frac{(c_W^2 m_{Z}^2 - m_{A'}^2) m_{A'}^2 \sigma^2}{m_{Z}^2 - m_{A'}^2} + \mathcal{O}(\sigma^3),
\end{align}
respectively. For $\sigma\sim 10^{-3}-10^{-2}$,  the relative mass shift is at most $0.3\%$. In later discussions, we will use $m_Z$ and $m_{A'}$ as the physical $Z$ boson mass and dark photon mass, respectively.

\section{Production and decay of dark photon}
\label{sec:prod_decay}

We now study the production and decays of $A'$.
Since we will consider $m_{A'}$ to be much larger than $m_Z$, therefore $A'$ can decay into fermion pairs, and also $W^+W^-$ and $Zh$ at the tree level. The couplings to fermion pairs, $W^+W^-$ and $Zh$ are given by
\begin{align}
& \mathcal{L}_{\text{fermions}} = \left (\epsilon e  Q_f\bar{f}\gamma^{\mu} f+  \tau \frac{g}{2c_W} \bar{f}\gamma^{\mu} (g_V^f-g_A^f\gamma_5) f \right )A_{\mu}^{\prime},\nonumber\\
& \mathcal{L}_{\text{gauge}} = -ie(\epsilon+\tau\cot\theta_W) \bigl[-\partial_{\mu}A_{\nu}^{\prime}(W_{\mu}^+W_{\nu}^--W_{\nu}^+W_{\mu}^-)\nn\\
&~~~~~~~~~~~~ +A_{\nu}^{\prime}(-W_{\mu}^+\partial_{\nu}W_{\mu}^-+W_{\mu}^-\partial_{\nu}W_{\mu}^+
+W_{\mu}^+\partial_{\mu}W_{\nu}^--W_{\mu}^-\partial_{\mu}W_{\nu}^+)\bigr],\nn\\
&\mathcal{L}_{\text{higgs}} = \dfrac{\tau g m_Z}{c_W}h A_{\mu}^{\prime}Z^{\mu}, 
\end{align}
where $f=e,\mu,\tau,\nu_{e,\mu,\tau},u,d,s,c,b$, $Q_f$ is the charge of the fermion, $g$ is the coupling constant of $SU(2)_L$, $g_V^f=T_3^f-2s_W^2 Q_f$ and $ g_A^f=T_3^f$ with $T_3= 1/2$, $-1/2$ for fermions being isospin $1/2$ and $-1/2$, respectively.

The cross section of $e^+e^-\to A^{\prime}\gamma$ is given by~\cite{Berends:1986yy}
\begin{align}
\label{eq:ap_production}
\sigma_{A^{\prime}\gamma}=-\dfrac{e^2(m_{A^{\prime}}^4+s^2)(1-\text{ln}\dfrac{s}{m_e^2})}{4\pi s^2(s-m_{A^{\prime}}^2)}\Big\{ e^2\epsilon^2 +\dfrac{g^2\tau^2\big[(g_V^{e})^2+(g_A^{e})^2\big]}{4c_W^2} -\dfrac{e g g^e_V \epsilon \tau}{c_W} \Big\},
\end{align}
where $g_V^e=-1/2+2s_W^2$, $g_A^e=-1/2$, and $\tau$ depends on $\sigma$ and $m_{A^{\prime}}$. 

\begin{figure}[!htb]
\centering
\includegraphics[width=0.4\textwidth]{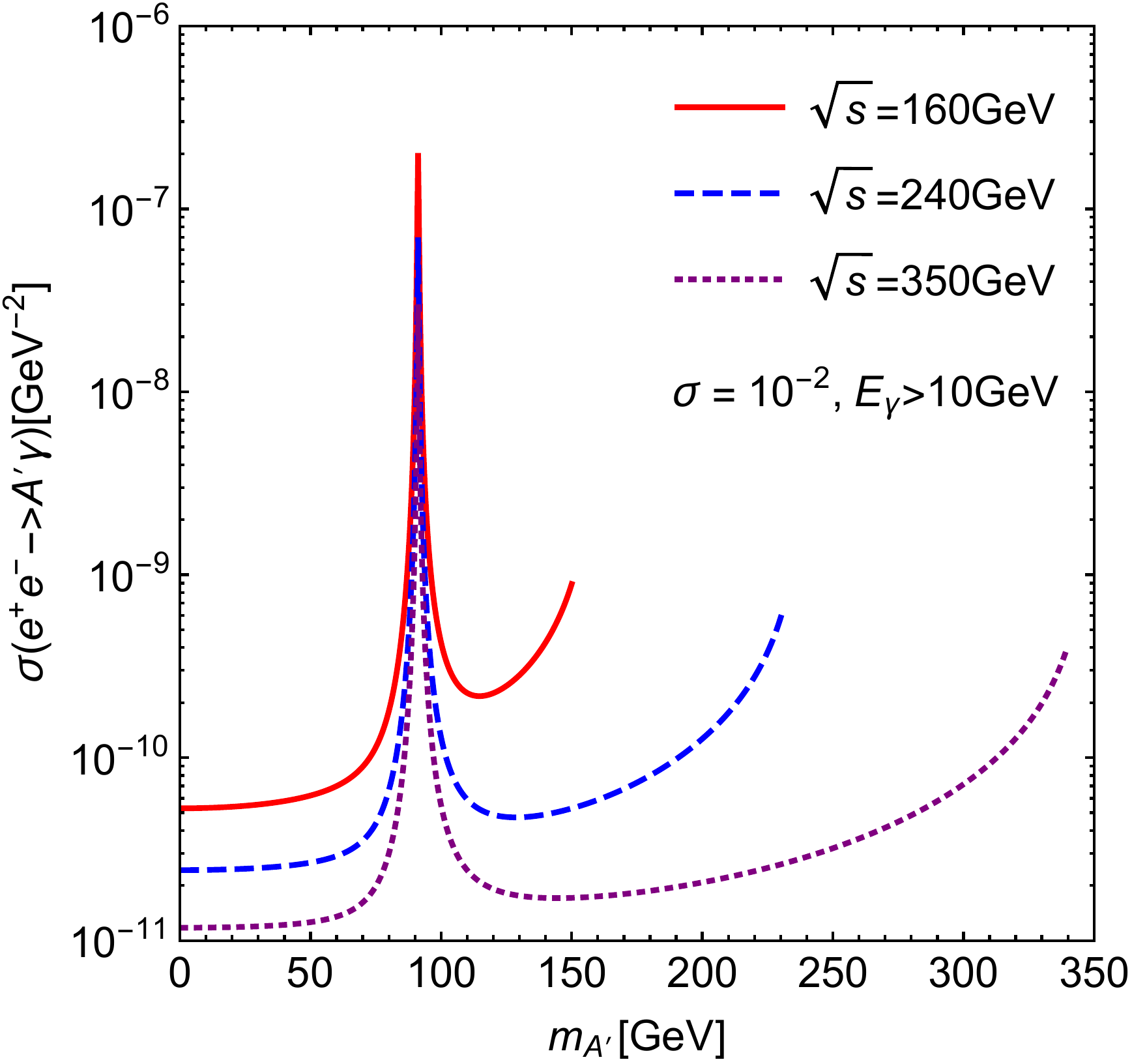}
\caption{Cross sections of $e^+ e^- \to A^\prime \gamma$ at $\sqrt{s}=160\gev$, $240\gev$ and $350\gev$ with $\sigma=10^{-2}$ and $E_{\gamma}>10\gev$.}
\label{fig:production}
\end{figure}

In Fig.~\ref{fig:production} we show the cross sections of $e^+e^-\to A^{\prime}\gamma$ at $\sqrt{s}=160\gev$, $240\gev$ and $350\gev$  for $\sigma=10^{-2}$. To keep the finiteness of the cross sections at $m_{A^{\prime}}=m_Z$, we use the complete forms of $\epsilon$ and $\tau$. However, the forms in Eq.~\eqref{coupling} are very good approximations as we have discussed. For $m_{A^{\prime}}$ being around $m_Z$ the cross sections exhibit peaks since the coupling $\tau$ is proportional to $m_{A^{\prime}}^2/(m_{A^{\prime}}^2-m_Z^2)$. Besides, the cross sections grow with $m_{A^{\prime}}$ for $m_{A^{\prime}}$ being above $m_Z$ due to $\sigma_{A^{\prime}\gamma}\sim {(m_{A^{\prime}}^4+s^2)}/{[s^2(s-m_{A^{\prime}}^2)]}$. However, if $m_{A^{\prime}}$ is very close to $\sqrt{s}$, that is the center-of-mass (c.m.) energy of $e^+e^-$ collider, the photon energy is small. In order to detect a isolated photon, a minimal energy of the photon should be demanded. In our case, the photon energy $E_{\gamma}>10\gev$ is required. 

Assuming that the dark photon $A^{\prime}$ decays exclusively to the SM particles, the partial widths of $A^{\prime}$ are~\cite{Barger:2009xg,Altarelli:1989ff}\footnote{There exist tri-boson decays $A^{\prime}\to W^+W^-\gamma$ and $A^{\prime}\to W^+W^-Z$, which originate from the SM quartic gauge couplings before the kinetic mixing. However, the tri-boson decay widths of the dark photon are smaller than $\mathcal{O}(1\%)$ so we neglect them in this paper.}
\begin{align}
\label{eq:decay_fermion}
\Gamma(A^{\prime}\to f\bar{f})&=\dfrac{g^2m_{A^{\prime}}}{12\pi c_W^2} N_c^f \bigl\{\epsilon^2 Q_f^2 c_W^2s_W^2 +\epsilon\tau Q_f c_W s_W g_V^f +\frac{1}{4}\tau^2 [(g_V^f)^2+(g_A^f)^2]\bigr\},\nn\\
\Gamma(A^{\prime}\to Zh)&=\dfrac{g^2\tau^2m_{A^{\prime}}}{192\pi c_W^2}\lambda^{1/2}(1,x_Z,x_h)\bigl\{\lambda(1,x_Z,x_h)+12x_Z\bigr\},\nn\\
\Gamma(A^{\prime}\to W^+W^-)&=\dfrac{g^2 s_W^2(\epsilon+\tau\cot\theta_W)^2m_{A^{\prime}}}{192\pi}x_{W}^{-2}(1-4x_W)^{3/2}(1+20x_W+12x_W^2),
\end{align}
where $N_c^f=3$ for quarks and 1 for leptons, $x_{W,Z,h}=(m_{W,Z,h}/m_{A^{\prime}})^2$ and $\lambda(x,y,z)=x^2+y^2+z^2-2xy-2yz-2zx$. The tree-level decay width into fermion pair in the first line of Eq.~\eqref{eq:decay_fermion} is a good approximation above the $b\bar{b}$ threshold $\Upsilon(nS)$, i.e., $m_{A^{\prime}}\gtrsim 12\gev$. The higher-order correction of $\Gamma(A^{\prime}\to f\bar{f})$ for $m_{A^{\prime}}\gtrsim 12\gev$ is small, for example the QCD correction to 3-loop level is $1.5\%$ for $m_{A^{\prime}}=60\gev$~\cite{limits3}, and not considered in this study.

\begin{figure}[!htb]
\centering
\includegraphics[width=0.4\textwidth]{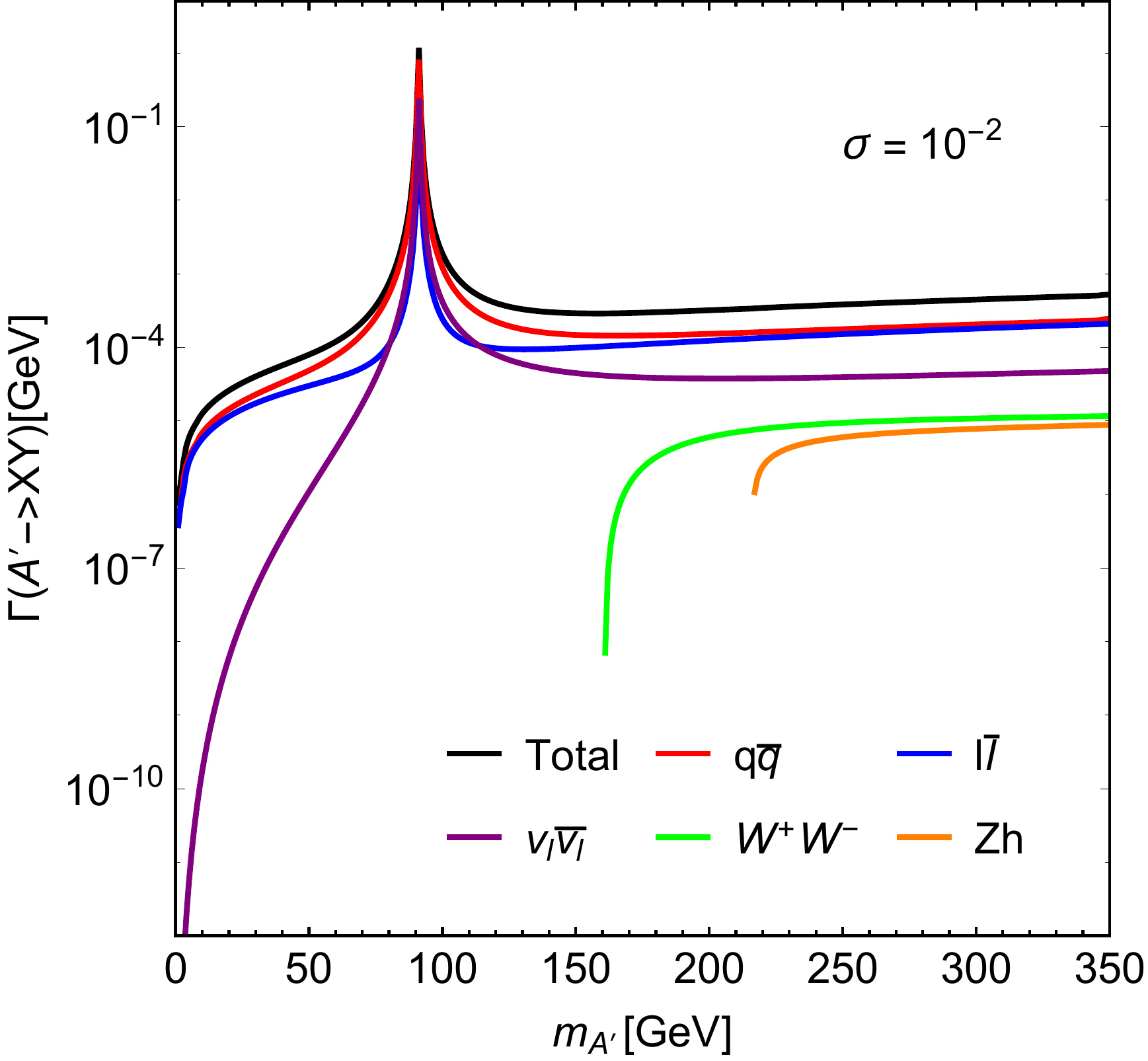}
\includegraphics[width=0.4\textwidth]{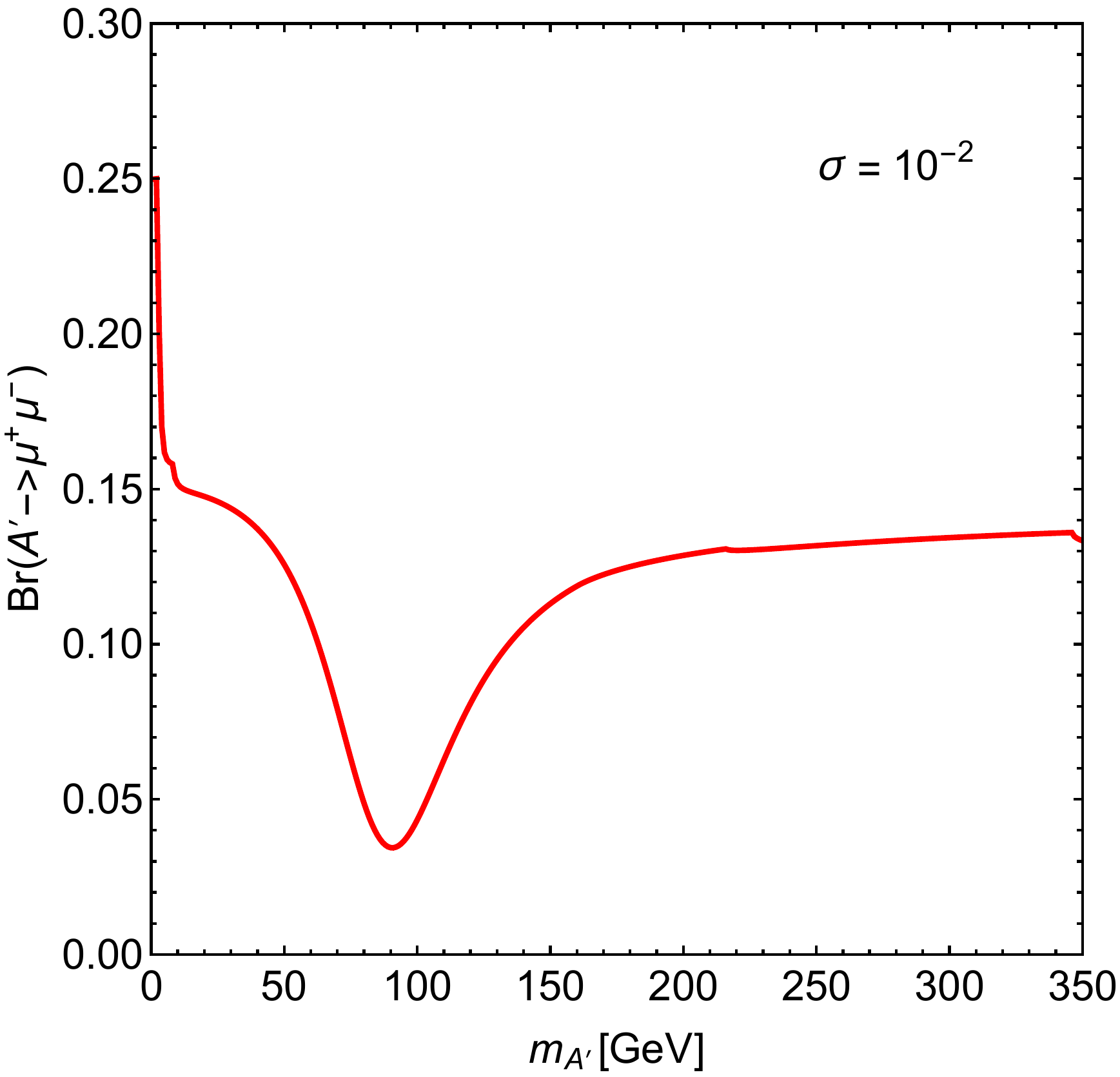}
\caption{Total and partial widths of $A^{\prime}$ (left) and the braching ratio of $A^{\prime}\to \mu^+\mu^-$ (right) with $\sigma=10^{-2}$. }
\label{fig:decaywidth}
\end{figure}

Figure~\ref{fig:decaywidth} displays the widths of $A^{\prime}$ and the braching ratio of $A^{\prime}\to \mu^+\mu^-$ in the left and right panels, respectively. Similar to the case of cross section, the complete forms of $\epsilon$ and $\tau$ are employed, so the branching ratio depends on $\sigma$ slightly. From the left panel, all the partial widths of $A^{\prime}$ into fermions exhibit peaks around $m_Z$ due to $\tau\sim m_{A^{\prime}}^2/(m_{A^{\prime}}^2-m_Z^2)$ and depend on $m_{A^{\prime}}$ linearly for $m_{A^{\prime}}\gtrsim 120\gev$ where the contributions from the second and third terms in the first line of Eq.~\eqref{eq:decay_fermion} are negligible. The decays into $W^+W^-$ and $Zh$ are possible once they are kinematically allowed. The widths $\Gamma(A^{\prime}\to Zh)$ and $\Gamma(A^{\prime}\to W^+W^-)$ are a few percent of the total width due to the suppression of phase space. Besides, the coupling of dark photon to $W^+W^-$ is proportional to $\epsilon+\tau\cot\theta_W=c_W\sigma m_{Z}^2/(m_{A^{\prime}}^2-m_Z^2)$, which becomes smaller for larger $m_{A^{\prime}}$. From the right panel, there is a dip around $m_Z$ due to the contribution from the coupling $\tau$. The branching ratio becomes flat for $m_{A^{\prime}}\gtrsim 150\gev$ where $\tau$ is small and the dependence on $m_{A^{\prime}}$ in $\Gamma(A^{\prime}\to \mu^+\mu^-)$ and the total width cancels.

\section{Sensitivities at future \tf{$e^+e^-$}{ee} colliders}
\label{sec:simulation}

In this section, we will study the sensitivities of searching for dark photon in the process $e^+e^-\to \gamma A^{\prime}, A^{\prime}\to \mu^+\mu^-$~\footnote{The $2\to 3$ process with off-shell dark photon and the interferene between the signal and background processes are checked to be small, which are thus neglected.} at the future $e^+e^-$ colliders with several different energies $\sqrt{s}=160\gev$, $240\gev$ and $350\gev$. The projected/updated integrated luminosities at the CEPC~\cite{CEPC-SPPCStudyGroup:2015csa}, FCC-ee~\cite{Gomez-Ceballos:2013zzn,Blondel:2017} and ILC~\cite{Baer:2013cma,Barklow:2015tja,Fujii:2017vwa} are summarized in Table~\ref{tab:collider_project}. At $\sqrt{s}=160\gev$, the integrated luminosity can be $10\abi$ at the FCC-ee with about 1-year running~\cite{Blondel:2017}, which is larger than that at the ILC. At $\sqrt{s}=240\sim 250\gev$, the integrated luminosity can be $5\abi$ at the CEPC and FCC-ee with 10-year~\cite{CEPC-SPPCStudyGroup:2015csa} and 3-year~\cite{Blondel:2017} running, respectively; while events with only $1.5\abi$ will accumulate at the ILC. At $\sqrt{s}=350\gev$, the integrated luminosity is projected to be $1.5\abi$ with $4\sim 5$-year running at the FCC-ee~\cite{Blondel:2017} and larger than that at the ILC. Thus in our study, we concentrate on the CEPC at $\sqrt{s}=240\gev$ and the FCC-ee with $\sqrt{s}=160\gev$ and $350\gev$.

\begin{table}[h]
\caption{The projected/updated integrated luminosities at the CEPC~\cite{CEPC-SPPCStudyGroup:2015csa}, FCC-ee~\cite{Gomez-Ceballos:2013zzn,Blondel:2017} and ILC~\cite{Baer:2013cma,Barklow:2015tja,Fujii:2017vwa}.}
\label{tab:collider_project}
\begin{center}
\begin{tabular}{ c | c | c | c }
\hline
Integrated luminosity $(\abi)$ & CEPC & FCC-ee & ILC   \\ \hline
$\sqrt{s}=160\gev$ & - & 10  &  0.5\\ \hline
$\sqrt{s}=240\sim 250\gev$ & 5  & 5  & 1.5 \\ \hline
$\sqrt{s}=350\gev$ & - & 1.5  & 0.2 \\ \hline
\hline
\end{tabular}
\end{center}
\end{table}

The most dominant SM background, for $e^+e^-\to \gamma A' \to \gamma \mu^+\mu^-$, is $e^+e^-\to \gamma (Z, \gamma)$ followed by virtual $Z$ and $\gamma$ decaying into $\mu^+\mu^-$. The analytic expressions of total cross sections and differential cross sections for the signal and background processes have been obtained in Ref.~\cite{He:2017ord}. In this work, we carry out numerical analyses with more realistic event selections at various future $e^+e^-$ colliders. For event generation, we use \texttt{MG5\_aMC\_v2\_4\_3}~\cite{Alwall:2014hca}~\footnote{Initial State Radiation (ISR) and beamstrahlung effect are not considered in this study.}. The following \textit{basic cuts} are imposed at the parton-level:
\begin{align}
|\eta_{\mu^\pm,\gamma}|<3,\quad E_{\gamma}>2\gev,\quad  \Delta R_{ij}>0.2,\quad \Delta m_{\mu^{+}\mu^{-}}< 10~\text{GeV},
\end{align}
where $\Delta R_{ij}=\sqrt{(\eta_i-\eta_j)^2+(\phi_i-\phi_j)^2}$ with $i,j=\mu^{\pm},\gamma$ and $\Delta m_{\mu^{+}\mu^{-}}\equiv | m_{\mu^{+}\mu^{-}}-m_{A^{\prime}}|$. The parton-level events are then interfaced with \texttt{Pythia6}~\cite{Sjostrand:2006za} for parton shower and hadronization. The detector effects are simulated with \texttt{Delphes-3.4.1}~\cite{deFavereau:2013fsa} and the built-in \texttt{delphes\_card\_CEPC.tcl} and \texttt{delphes\_card\_ILD.tcl} for the CEPC and FCC-ee, respectively. The detector parametrization for muon momentum resolution and electromagnetic calorimeter (ECAL) energy resolution are
\begin{itemize}
\item[-] $\dfrac{\Delta p_T}{p_T}=0.1\%\oplus \dfrac{p_T}{10^5 \gev}$ for $|\eta|<1.0$ and 10 times larger for $1.0<|\eta|<3.0$;
\item[-] $\dfrac{\Delta E}{E}=\dfrac{0.20}{\sqrt{E/\text{GeV}}}\oplus 0.5\%.$ for $|\eta|<3.0$.
\end{itemize}
for the CEPC and 
\begin{itemize}
\item[-] $\dfrac{\Delta p_T}{p_T}=0.1\%\oplus \dfrac{p_T}{10^5 \gev}$ for $|\eta|<1.0$ and 10 times larger for $1.0<|\eta|<2.4$;
\item[-] $\dfrac{\Delta E}{E}=\dfrac{0.15}{\sqrt{E/\text{GeV}}}\oplus 1\%.$ for $|\eta|<3.0$.
\end{itemize}
for the FCC-ee.

In order to identify objects in the final state, we impose the following \textit{pre-selection criteria}: 
\begin{itemize}
\item A pair of opposite-sign muons are selected with $E_{\mu^{\pm}}>2\gev$~\cite{Yu:2017mpx}, $|\eta_{\mu^\pm}|<2.5$~\cite{CEPC-SPPCStudyGroup:2015csa,Chen:2016zpw} for the CEPC with a better muon identification than the ILD-like detector performance: $p_{T\mu^{\pm}}>10\gev$~\cite{Cerri:2016bew}, $|\eta_{\mu^\pm}|<2.5$~\cite{Abe:2010aa}, which is used for the simulation at the FCC-ee;
\item Exactly one photon with $p_{T}^{\gamma}\geq 10\gev$~\cite{Abe:2010aa,Cerri:2016bew} and $|\eta_{\gamma}|<2.5$~\cite{CEPC-SPPCStudyGroup:2015csa} is selected;
\item $\Delta R_{ij}>0.4$ for $i,j=\mu^{\pm},\gamma$.
\end{itemize}

\begin{figure}[!htb]
\centering
\includegraphics[width=0.3\textwidth]{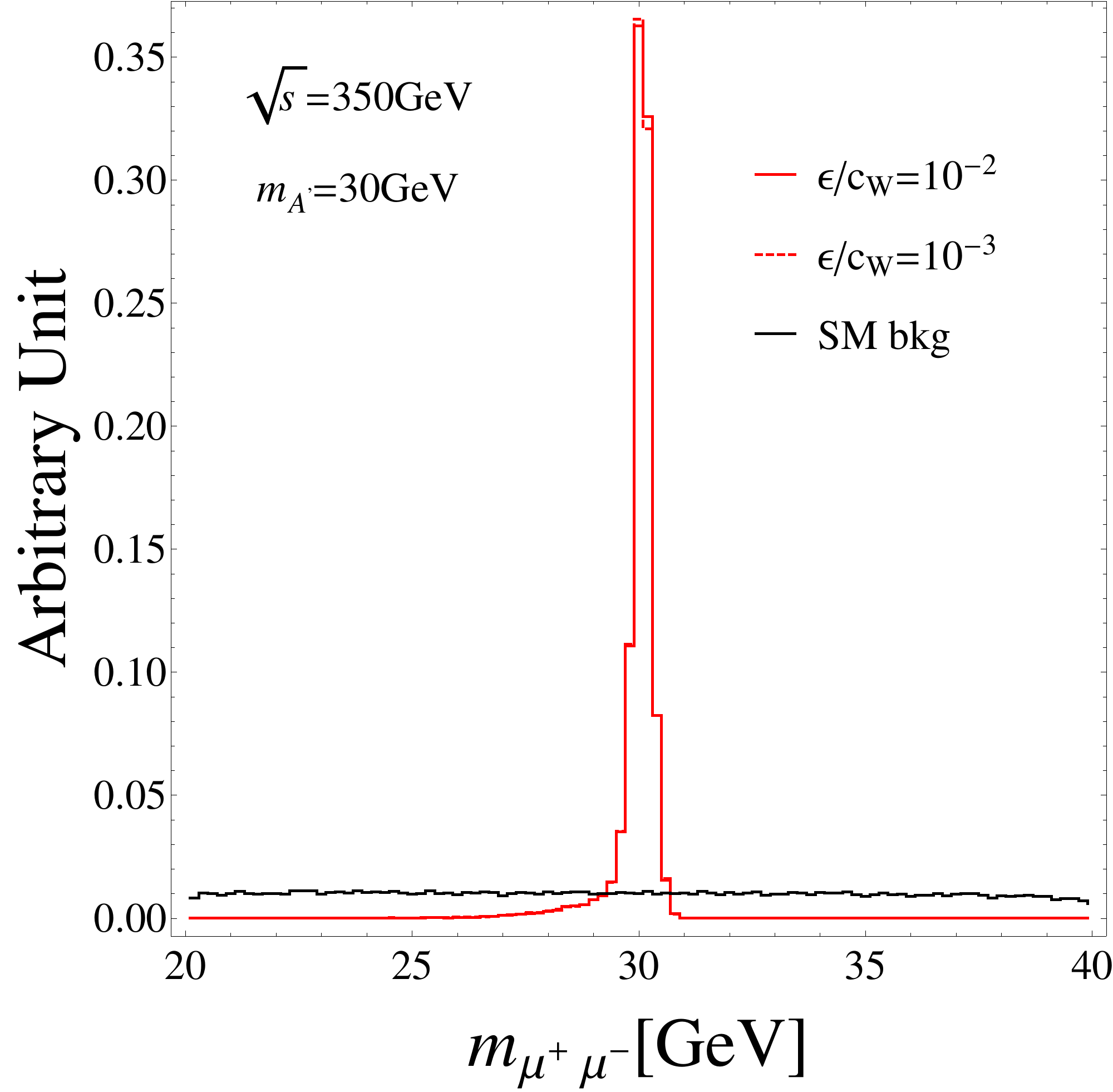}
\includegraphics[width=0.3\textwidth]{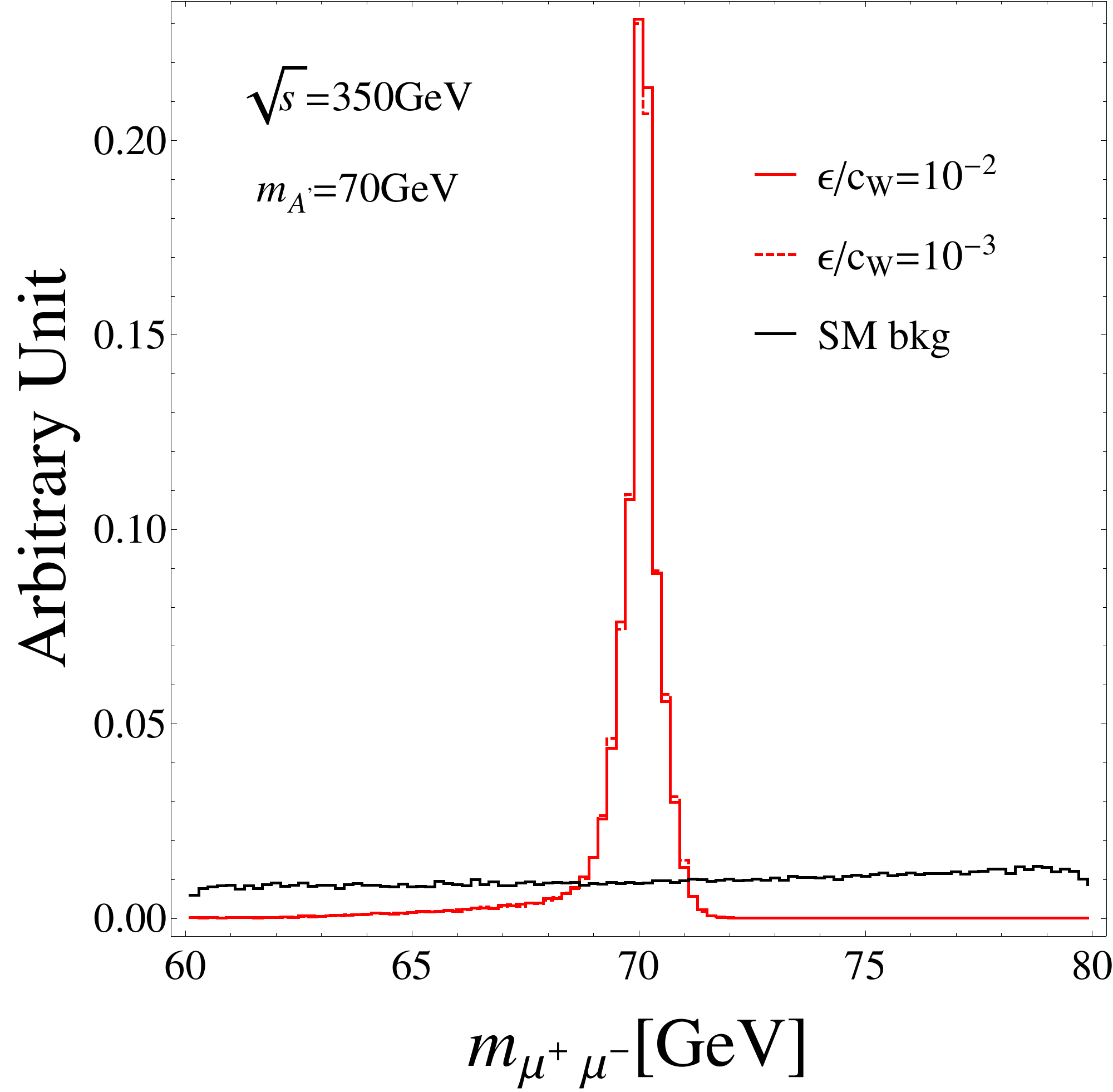}
\includegraphics[width=0.3\textwidth]{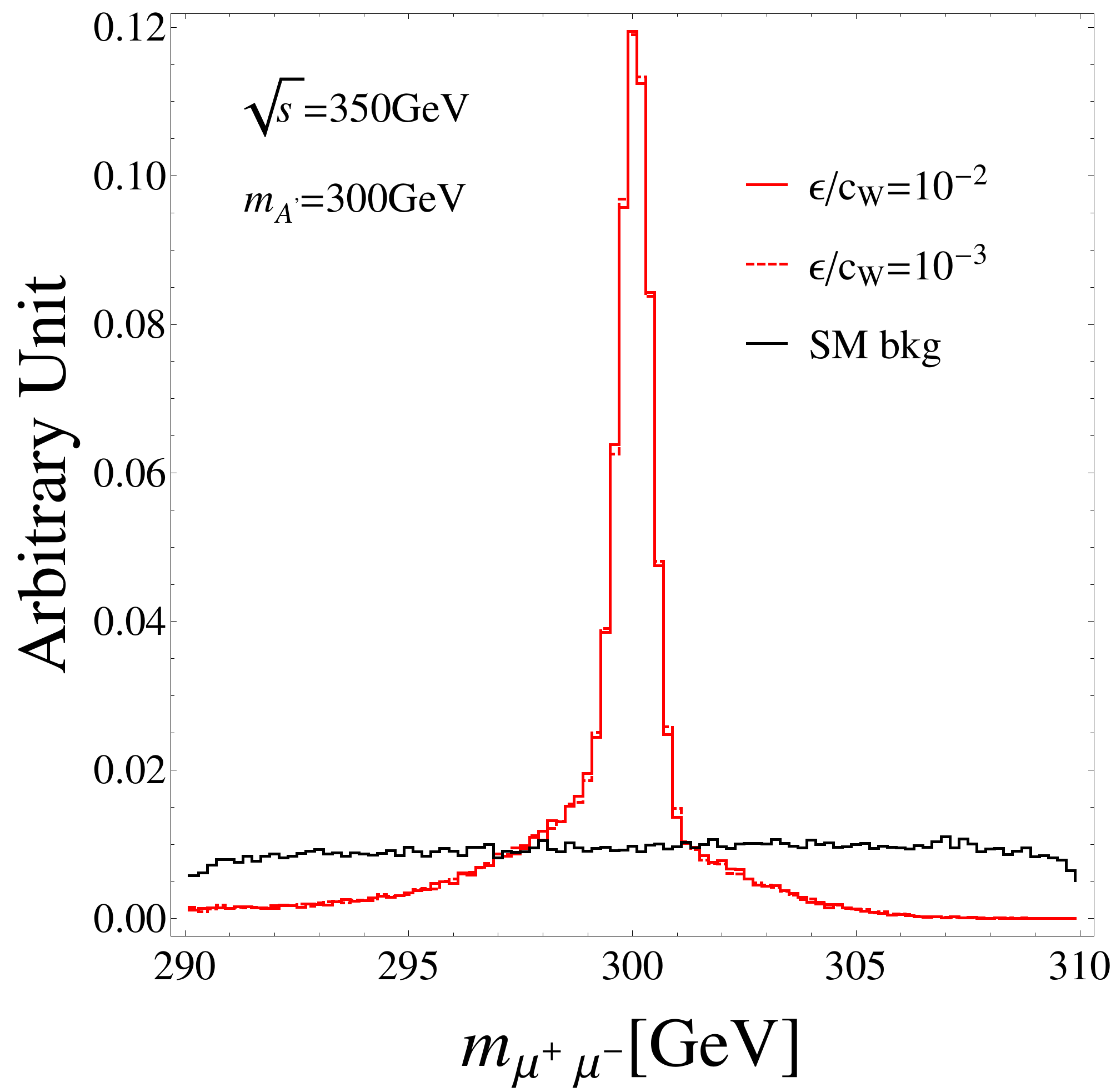}
\caption{The normalized $m_{\mu^+\mu^-}$ distributions for $m_{A^{\prime}}=30\gev$, $70\gev$, $300\gev$ with $\epsilon/c_W=10^{-2}$ and $\epsilon/c_W=10^{-3}$ at $\sqrt{s}=350\gev$. The distributions are not sensitive to the total width and thus $\epsilon$ for $\epsilon/c_W\sim 10^{-3}-10^{-2}$.}
\label{fig:kinematics_FCCee}
\end{figure}

After passing these pre-selection criteria, the invariant mass distributions of $\mu^+\mu^-$ are displayed in Fig.~\ref{fig:kinematics_FCCee}, where we have chosen three benchmark values of $m_{A^{\prime}}$, i.e.,  $m_{A^{\prime}}=30\gev$, $70\gev$, $300\gev$, with $\epsilon/c_W=10^{-2}$ and $10^{-3}$ at the FCC-ee with $\sqrt{s}=350\gev$ (the results are similar at the CEPC with $\sqrt{s}=240\gev$ and the FCC-ee with $\sqrt{s}=160\gev$). The invariant mass distributions of $\mu^+\mu^-$ are affected by the muon momentum resolution and the total width of dark photon, which is proportional to $\epsilon^2$. Figure~\ref{fig:kinematics_FCCee} shows that for $\epsilon/c_W\sim 10^{-3}-10^{-2}$ the impact of the dark photon total width on the invariant mass distributions is negligible. Thus it is flexible to choose $\epsilon/c_W=10^{-2}$~\footnote{Actually in the very small region of $m_{A^{\prime}}\simeq m_Z$, the total width of dark photon becomes significantly large, thus the cut efficiency of the signal depends on the value of $\epsilon$.} in our collider study. 

The energy of dark photon can be expressed as
\begin{align}
E_{A^{\prime}}=\dfrac{\sqrt{s}}{2}(1+\dfrac{m_{A^{\prime}}^2}{s}),
\end{align}
where $\sqrt{s}$ is the c.m. energy of the collider. Thus the invariant mass distribution becomes broader with larger $m_{A^{\prime}}$ or the increase of $\sqrt{s}$ from $160\gev$ to $350\gev$. 

Based on the kinematical distributions, we further impose the \textit{selection cuts}:
\begin{align}
\Delta m_{\mu^{+}\mu^{-}} < 0.5\sim 1.5~\text{GeV},\quad E_T^{\text{miss}}< 5\gev,
\end{align}
where the $\Delta m_{\mu^{+}\mu^{-}}$ cut is explicitly shown in Table~\ref{tab:deltam}. The missing transverse momentum $(E_T^{\text{miss}})$ cut is used to remove the SM backgrounds $\tau^+\tau^-\gamma$ and $W^+W^-\gamma$, which have larger $E_T^{\text{miss}}$~\footnote{There is also SM background from $e^+e^-\to h\gamma, h\to \mu^+\mu^-$. We have checked that it is small and can be neglected.}.

On the other hand, in Ref.~\cite{Liu:2017lpo} the following selection cuts were imposed:
\begin{align}
\Delta m_{\ell^+\ell^-}\equiv |m_{\ell^+\ell^-}-m_{A^{\prime}}| < 5~\text{GeV},\quad |E_{\gamma}-\frac{s-m_{A^{\prime}}^2}{2\sqrt{s}}|<2.5\gev.
\end{align}
We find that after our mass window cut $\Delta m_{\mu^{+}\mu^{-}} < 0.5\sim 1.5~\text{GeV}$ is imposed, the above cut on the photon energy spectrum is not effective. In Ref.~\cite{Karliner:2015tga}, the total and differential cross sections of $e^+e^-\to \gamma A^{\prime}\to \gamma\mu^+\mu^-$ were expressed as those of $e^+e^-\to A^{\prime}\to \mu^+\mu^-$ convoluted with the probability function of the emitted photon~\cite{Chen:1975sh}. For signal extraction, an estimated mass resolution $\Delta m_{\mu^+\mu^-} = m_{A^{\prime}}^2/(10^5\gev)$ was adopted~\cite{Karliner:2015tga}, which was based on the specification of the muon momentum resolution $\Delta(1/p_T)=2\times 10^{-5}\gev^{-1}$. This estimation, in our opinion, is too optimistic especially for low $p_T$ muons.

\begin{table}[h]
\caption{The dependence of $\Delta m_{\mu^{+}\mu^{-}}$ on $m_{A^{\prime}}$ and the c.m. energy. FCC-ee (160~GeV) and FCC-ee (350~GeV) denote the FCC-ee at $\sqrt{s}=160~\gev$ and $350~\gev$, respectively. CEPC (240~GeV) denotes the CEPC at $\sqrt{s}=240~\gev$. All the numbers in the table are in units of $\gev$. }
\label{tab:deltam}
\begin{center}
\begin{tabular}{ c | c | c | c }
\hline
$m_{A^{\prime}}$ & FCC-ee (160~GeV) & CEPC (240~GeV) & FCC-ee (350~GeV)   \\ \hline
$[20,40]$ & 0.5 & 0.5  & 0.5 \\ \hline
$[50,60]$   & 0.5 & 1.0  & 1.0 \\ \hline
$[70,94]$ & 1.0 & 1.0  & 1.0 \\ \hline
$\geq 95$ & 1.5 & 1.5 & 1.5 \\ \hline
\hline
\end{tabular}
\end{center}
\end{table}

\begin{figure}[!htb]
\centering
\includegraphics[width=0.5\textwidth]{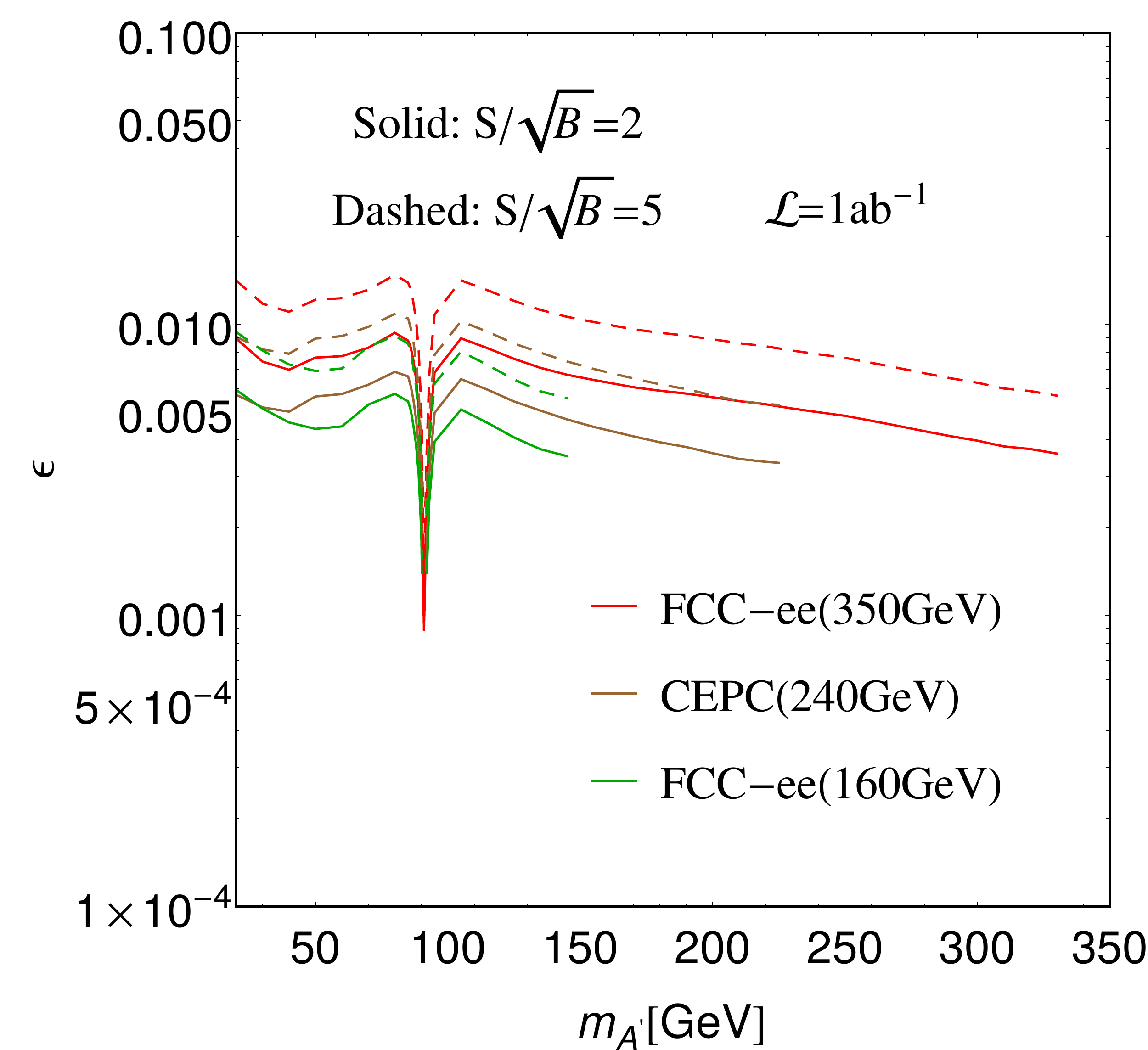}
\caption{Sensitivities to $\epsilon$ at the FCC-ee (160~GeV), CEPC (240~GeV) and FCC-ee (350~GeV) with the same integrated luminosity $\mathcal{L}=1\abi$.}
\label{fig:sensitivity_CEPC}
\end{figure}

The signal significance is evaluated using
\begin{align}
\label{eq:sensitivity}
\frac{S}{\sqrt{B}}&=(\frac{S}{\sqrt{B}})_0 \dfrac{\epsilon^2}{10^{-4}}\sqrt{\dfrac{\mathcal{L}}{1\abi}},
\end{align}
where the benchmark value of the significance $(S/\sqrt{B})_0$ is evaluated with $\epsilon^2=10^{-4}$ and $\mathcal{L}=1\abi$.

Figure~\ref{fig:sensitivity_CEPC} shows the sensitivities to $\epsilon$ with $S/\sqrt{B}=2$ and $S/\sqrt{B}=5$ at the FCC-ee (160~GeV), CEPC (240~GeV) and FCC-ee (350~GeV) with the same integrated luminosity $\mathcal{L}=1\abi$. We find that  FCC-ee (160~GeV) has the best sensitivity for $20\gev<m_{A^{\prime}}<140\gev$ as compared to the CEPC (240~GeV) and FCC-ee (350~GeV) and the sensitivities increase for larger $m_{A^{\prime}}$ if $m_{A^{\prime}}\gtrsim 120\gev$. This is mainly because that the cross section of $e^+e^-\to A^{\prime}\gamma$ at $\sqrt{s}=160\gev$ is largest and the cross sections at $\sqrt{s}=160\gev$, $240\gev$ and $350\gev$ increase for $m_{A^{\prime}}\gtrsim 120\gev$ as shown in Fig.~\ref{fig:production}. Note that the sensitivities are only shown with $m_{A^{\prime}}\lesssim \sqrt{s}-20\gev$ beyond which the signal events are hard to pass the selection $p_T^{\gamma}\geq 10\gev$.

As discussed in section~\ref{sec:intro}, the LHC Drell-Yan process can provide constraints for dark photon mass above 10~GeV. For the dark photon mass $10\gev<m_{A^{\prime}}<80\gev$, the sensitivities to $\epsilon$ has been explored~\cite{Hoenig:2014dsa} by recasting the CMS 7 TeV measurements~\cite{Chatrchyan:2013tia} and making projects for the 8 TeV LHC with the integrated luminosity $\mathcal{L}=20~\text{fb}^{-1}$ and at the High-Luminosity LHC (HL-LHC) with $\sqrt{s}=14~\text{TeV}$ and $\mathcal{L}=3\abi$. The limits on $\epsilon$ for $m_{A^{\prime}}\gtrsim 180\gev$ were derived~\cite{Cline:2014dwa} using the ATLAS 8~TeV measurements with $\mathcal{L}=20~\text{fb}^{-1}$~\cite{ATLAS:2013jma} (the published version is Ref.~\cite{Aad:2014cka}). The results were then projected to the HL-LHC in Ref.~\cite{limits3}. In this work, we recast the 13 TeV measurents at high mass region $(150\gev\sim 350\gev)$ with $\mathcal{L}=36.1\fbi$~\cite{Aaboud:2017buh} and project it to the sensitivities at $300\fbi$ and $3\abi$. Specifically, we generate the leading-order (LO) process $pp\to A^{\prime}\to \mu^+\mu^-$ in \texttt{MG5\_aMC\_v2\_4\_3} with $\epsilon/c_W=10^{-2}$ and obtain the cross section of $pp\to A^{\prime}\to \mu^+\mu^-$, denoted as $\sigma_{\text{LO}}(A^{\prime})\text{Br}(\mu^+\mu^-)$, for $m_{A^{\prime}}$ in the range $150\gev \sim 350\gev$. The LO cross section is then multiplied with a next-to-leading-order (NLO) QCD $K$-factor $K_{\text{NLO}}\simeq 1.2$~\cite{Fuks:2017vtl}. The experimental 95\% confidence level (C.L.) upper limit on the cross section times the branching ratio of $pp\to A^{\prime}\to \mu^+\mu^-$ with $36.1\fbi$, denoted as $[\sigma(A^{\prime})\text{Br}(\mu^+\mu^-)]^{95\%\text{C.L.}}$, is extracted from Ref.~\cite{Aaboud:2017buh} directly and independent of $\epsilon$. The 95\% C.L. upper limit on $\epsilon$ with the integrated luminosity $\mathcal{L}$ is thus
\begin{align}
\epsilon^{95\% \text{C.L.}}= \bigg( \dfrac{[\sigma(A^{\prime})\text{Br}(\mu^+\mu^-)]^{95\%\text{C.L.}}}{K_{\text{NLO}}\sigma_{\text{LO}}(A^{\prime})\text{Br}(\mu^+\mu^-)/(10^{-4}c_W^2)}\sqrt{\dfrac{36.1\fbi}{\mathcal{L}}}  \bigg)^{1/2}.
\end{align} 

Besides the direct searches in the Drell-Yan process, the mixing between $Z_0$ and $A_{0}^{\prime}$ leads to shifts in the mass of $Z$ boson and its couplings to the SM fermions, which are confronted with the electroweak precision tests (EWPTs)~\cite{limits3,Hook:2010tw}.

\begin{figure}[!htb]
\centering
\includegraphics[width=0.5\textwidth]{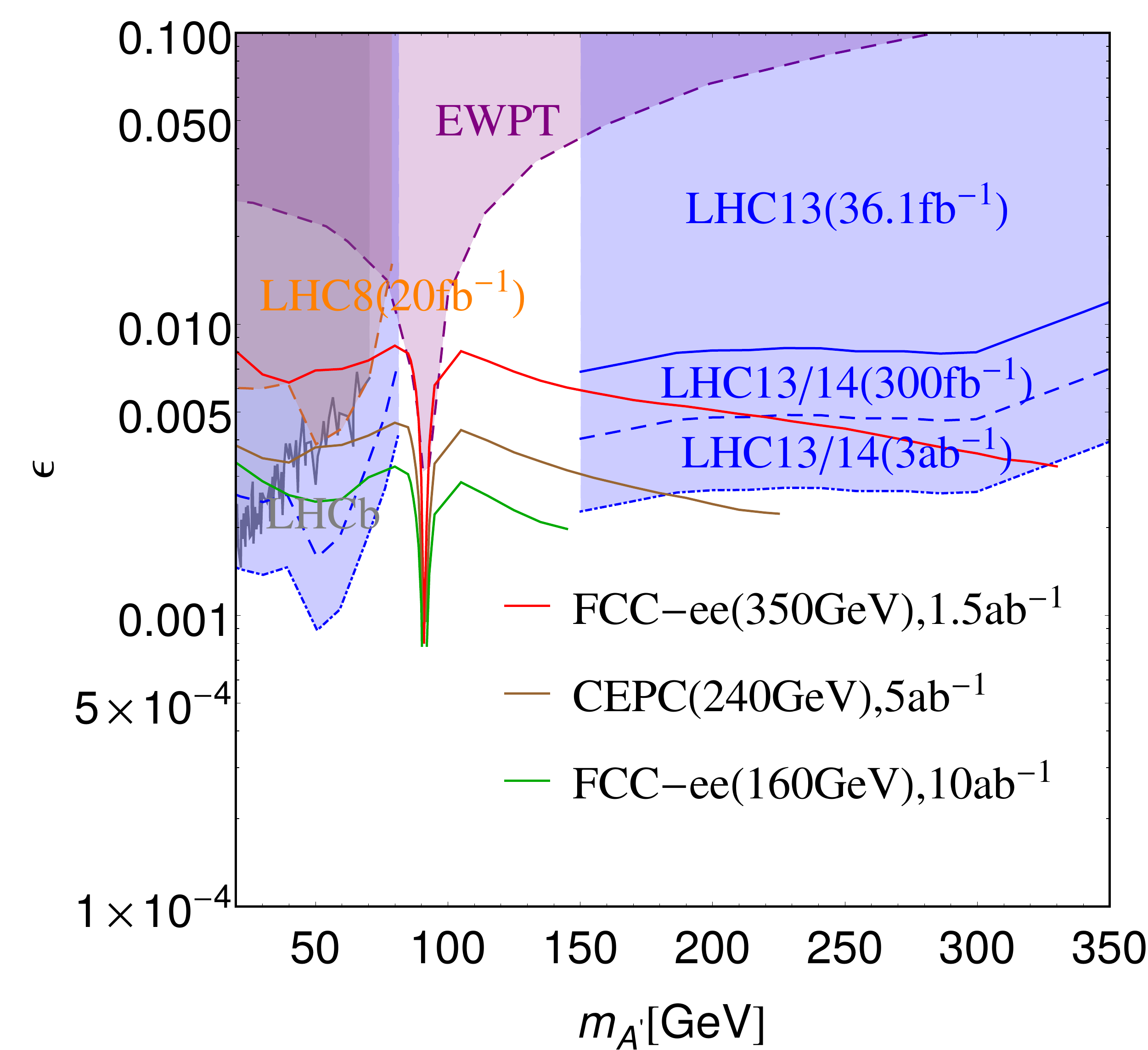}
\caption{Exclusion limits to $\epsilon$ at different colliders. The sensitivities with $S/\sqrt{B}=2$ at the CEPC with $\mathcal{L}=5\abi$ (brown curve) and at the FCC-ee with $\mathcal{L}=10\abi$ (green curve), $1.5\abi$ (red curve) for $\sqrt{s}=160\gev$, $350\gev$ are shown. The 95\% C.L. constraints from the EWPTs (purple region), LHC 8 TeV Drell-Yan process with $\mathcal{L}=20\fbi$ (orange region),  LHC 14 TeV Drell-Yan process with $\mathcal{L}=3\abi$ for $10\gev<m_{A^{\prime}}<80\gev$ (blue region) are taken from Ref.~\cite{limits3} (in fact, QCD $K$-facotrs should also be included for recasting the results in the Drell-Yan process at the LHC, which will lead to stricter constraints), which is also rescaled to that with $\mathcal{L}=300\fbi$ (blue dashed region). The 90\% C.L. constraint from the LHCb prompt search is taken from Ref.~\cite{Aaij:2017rft}. The 95\% C.L. constraints for $m_{A^{\prime}}\gtrsim 150\gev$ at the 13 TeV LHC with $\mathcal{L}=36.1\fbi$ (blue region), $300\fbi$ (blue dashed region) and $3\abi$ (blue dot-dashed region) are also shown. }
\label{fig:final}
\end{figure}

The exclusion limits at future $e^+e^-$ colliders and those from direct searches in the Drell-Yan process $pp\to A^{\prime}\to \ell^+\ell^-$ at the LHC, LHCb prompt searches~\cite{Aaij:2017rft} and the EWPTs are illustrated in Fig.~\ref{fig:final} for the dark photon mass $20\gev<m_{A^{\prime}}\lesssim 340\gev$. For the dark photon mass below $70\gev$, the LHCb prompt searches have imposed severe constraints on $\epsilon$, which is $\epsilon\lesssim 2\times 10^{-3}-6\times 10^{-3}$. The  FCC-ee (160~GeV) and CEPC (240~GeV) can have better sensitivity with $m_{A^{\prime}}\gtrsim 50\gev$. There does not exist constraints on $\epsilon$ from the LHC direct searches in the Drell-Yan process for $80\gev\lesssim m_{A^{\prime}}\lesssim 150\gev$, while the constraint from the EWPTs is weak for $m_{A^{\prime}}\gtrsim 100\gev$. The exclusion limit can be improved significantly at future $e^+e^-$ colliders, which can even reach $2\times 10^{-3}$ at the FCC-ee (160~GeV) with $\mathcal{L}=10\abi$. For larger dark photon mass $150\gev\lesssim m_{A^{\prime}}\lesssim 300\gev$, the current constraint from the LHC direct searches is $\epsilon\lesssim 8.3\times 10^{-3}$, which is projected to be $\epsilon\lesssim 4.8\times 10^{-3}$ and $\epsilon\lesssim 2.7\times 10^{-3}$ with the integrated luminosity $\mathcal{L}=300\fbi$ and $3\abi$, respectively. The sensitivity at the FCC-ee (350~GeV) with $1.5\abi$ is better than that at the 13~TeV LHC with $300\fbi$ for $m_{A^{\prime}}\gtrsim 220\gev$. While the sensitivity at the CEPC (240~GeV) with $5\abi$ can be even better than that at 13~TeV LHC with $3\abi$ for $m_{A^{\prime}}\gtrsim 180\gev$. For the dark photon mass larger than about 300~GeV, the sensitivity to $\epsilon$ may be further improved at a $e^+e^-$ collider with larger c.m. energy.

\section{Summary and conclusions}
\label{sec:summary}

In this work we study dark photon search using $e^+e^- \to \gamma A' \to \gamma \mu^+ \mu^-$ for a dark photon mass $m_{A'}$ as large as kinematically allowed at future $e^+e^-$ colliders. For small dark photon mass, the mixing is small for a small mixing parameter $\sigma$. For large $m_{A'}$, care should be taken to properly treat possible large mixing between $A'$ and $Z$. We show that stringent constraints on the parameter $\epsilon$ for a wide range of dark photon mass can be obtained at planed $e^+\;e^-$ colliders, such as CEPC, ILC and FCC-ee.

As compared to previous studies with the estimated mass resolution $\Delta m_{\mu^+\mu^-} = m_{A^{\prime}}^2/(10^5\gev)$~\cite{Karliner:2015tga} at $\sqrt{s}=90\gev$ and $250\gev$ or the mass window cut $\Delta m_{\mu^+\mu^-}<5\gev$~\cite{Liu:2017lpo} at $\sqrt{s}=250\gev$ and $500\gev$, our study with a detailed detector simulation and a realistic muon momentum resolution shows that a mass window cut $\Delta m_{\mu^+\mu^-}<0.5\gev\sim 1.5\gev$ is appropriate for the dark photon searches in $e^+e^-\to \gamma A^{\prime}\to \gamma\mu^+\mu^-$ at future $e^+e^-$ colliders with $\sqrt{s}=160\gev$, $240~(250)\gev$ and $350\gev$. Consequently, our results are optimized and more realistic. Epecifically, we find that the $2\sigma$ exclusion limits on $\epsilon$ for the dark photon from 20~GeV to 330~GeV  are $\epsilon\lesssim 10^{-3}-10^{-2}$ at future $e^+e^-$ colliders.
The CEPC (240~GeV) and FCC-ee (160~GeV) are more sensitive than the constraint from current LHCb measurement once the dark photon mass $m_{A^{\prime}}\gtrsim 50\gev$. We also obtain the constraint on $\epsilon$ for $150\gev\lesssim m_{A^{\prime}}\lesssim 350\gev$ from the direct searches in the Drell-Yan process $pp\to X\mu^+\mu^-$ using the 13 TeV LHC measurements with $\mathcal{L}=36.1\fbi$~\cite{Aaboud:2017buh} and project it to the measurements with $\mathcal{L}=300\fbi$ and $3\abi$. The corresponding constraints are $\epsilon\lesssim 8.3\times 10^{-3}$, $\epsilon\lesssim 4.8\times 10^{-3}$ and $\epsilon\lesssim 2.7\times 10^{-3}$ for $150\gev\lesssim m_{A^{\prime}}\lesssim 300\gev$, and become weaker for $m_{A^{\prime}}\gtrsim 300\gev$. For $m_{A^{\prime}}\gtrsim 220\gev$, the sensitivity at the FCC-ee (350~GeV) with $1.5\abi$ is better than that at the 13~TeV LHC with $300\fbi$, while the sensitivity at the CEPC (240~GeV) with $5\abi$ can be even better than that at 13~TeV LHC with $3\abi$ for $m_{A^{\prime}}\gtrsim 180\gev$. Besides, we have compared the sensitivities at $\sqrt{s}=160\gev$, $240\gev$ and $350\gev$ with the same integrated luminosity and find that the sensitivity at the $160\gev$ $e^+e^-$ collider is better than the other two due to its largest cross section for on-shell dark photon production.

\acknowledgments
We would like to thank Qing-Hong Cao, Gang Li (IHEP), Tanmoy Modak, Manqi Ruan and Hao Zhang for valuable discussions. This work was supported in part by the MOST (Grant No. MOST104-2112-M-002-015-MY3 and 106-2112-M-002-003-MY3 ), and in part by Key Laboratory for Particle Physics,
Astrophysics and Cosmology, Ministry of Education, and Shanghai Key Laboratory for Particle
Physics and Cosmology (Grant No. 15DZ2272100), and in part by the NSFC (Grant Nos. 11575111 and 11735010).

\appendix

%
\end{document}